\documentclass[aps,prl,twocolumn,superscriptaddress]{revtex4-1}
\usepackage{graphicx}
\usepackage{amsmath}
\usepackage{amssymb}
\usepackage{colordvi}
\usepackage{mathrsfs}
\usepackage{bm}
\usepackage{verbatim}
\usepackage{dcolumn}
\usepackage{epsfig}
\usepackage{subfigure}
\usepackage{makecell}
\usepackage[colorlinks,linkcolor=blue,anchorcolor=blue,citecolor=blue]{hyperref}

\begin{document}
\title{Berry-Curvature Engineering for Nonreciprocal Directional Dichroism in Two-Dimensional Antiferromagnets}

\author{Wenhao Liang}
\affiliation{International Centre for Quantum Design of Functional Materials, CAS Key Laboratory of Strongly-Coupled Quantum Matter Physics, and Department of Physics, University of Science and Technology of China, Hefei, Anhui 230026, China}

\author{Junjie Zeng}
\affiliation{Institute for Structure and Function \& Department of Physics \& Chongqing Key Laboratory for Strongly Coupled Physics, Chongqing University, Chongqing 400044, P. R. China}

\author{Zhenhua Qiao} \email[Correspondence author:~~]{qiao@ustc.edu.cn}
\affiliation{International Centre for Quantum Design of Functional Materials, CAS Key Laboratory of Strongly-Coupled Quantum Matter Physics, and Department of Physics, University of Science and Technology of China, Hefei, Anhui 230026, China}
\affiliation{Hefei National Laboratory, University of Science and Technology of China, Hefei 230088, China}

\author{Yang Gao}\email[Correspondence author:~~]{ygao87@ustc.edu.cn}
\affiliation{International Centre for Quantum Design of Functional Materials, CAS Key Laboratory of Strongly-Coupled Quantum Matter Physics, and Department of Physics, University of Science and Technology of China, Hefei, Anhui 230026, China}
\affiliation{Hefei National Laboratory, University of Science and Technology of China, Hefei 230088, China}

\author{Qian Niu}
\affiliation{International Centre for Quantum Design of Functional Materials, CAS Key Laboratory of Strongly-Coupled Quantum Matter Physics, and Department of Physics, University of Science and Technology of China, Hefei, Anhui 230026, China}

\date{\today{}}

\begin{abstract}
  In two-dimensional antiferromagnets, we identify the mixed Berry curvature as the geometrical origin of the nonreciprocal directional dichroism (NDD), which refers to the difference in light absorption with the propagation direction flipped. Such a Berry curvature is strongly tied to the uniaxial strain in accordance with the symmetry constraint, leading to a highly tunable NDD, whose sign and magnitude can be manipulated via the strain direction. 
  As a concrete example, we demonstrate such a phenomenon
  in a lattice model of MnBi$_2$Te$_4$. The coupling between the mixed Berry curvature and strain also suggests the magnetic quadrupole of the Bloch wave packet as the macroscopic order parameter probed by the NDD in two dimensions, distinct from the multiferroic order $\bm P \times \bm M$ or the spin toroidal and quadrupole order within a unit cell in previous studies. Our work paves the way of the Berry-curvature engineering for optical nonreciprocity in two-dimensional antiferromagnets.
\end{abstract}

\maketitle

Antiferromagnetism in two dimensions has attracted growing interest in recent years~\cite{AFM-4, AFM-5, AFM-6, AFM-7, AFM-review-1, AFM-review-2, AFM-review-3, AFM-review-4}.
Benefiting from both the ultrafast dynamics and absence of stray fields from antiferromagnetism~\cite{AFM-review-1}, and remarkable tunability from two-dimensional van der Waals structures~\cite{2Dmaterials-1, 2Dmaterials-2, 2Dmaterials-3, 2Dmaterials-4}, two-dimensional antiferromagnets can play an essential role in designing next-generation spintronic devices~\cite{AFM-review-1, AFM-review-2, AFM-review-3, AFM-review-4}.
Their connection to topological and geometrical physics is also recognized~\cite{AFMtopo-2, AFMtopo-3, AFMtopo-4, MBT-liang,AFMBC-1, AFMBC-2, AFMBC-3, LHE}. This naturally integrates spintronics with topological protection as well as distinct tuning scheme via Berry-curvature engineering. One fundamental and timely question to answer in this regard is how the geometrical quantity (e.g., Berry curvature) is intertwined with various degrees of freedom, such as charge, spin, and lattice. One such example is the layer Hall effect in the antiferromagnetic MnBi$_2$Te$_4$ system due to the locking of Berry curvature with different layers~\cite{LHE}. Hereinbelow, we investigate the coupling between the Berry curvature and uniaxial strain through the nonreciprocal directional dichroism~(NDD).

NDD refers to the difference in optical absorption between counter-propagating lights with linear polarization~\cite{NDD-1965, NDD-gao, NDD-first Xray, NDD-Xray, NDD-one way, NDD-nagaosa, NDD-enhanced,  NDD-skyrmion, ME-electromagnon-1, NDD-ME-1, NDD-ME-2, NDD-ME-3, NDD-T-1, NDD-T-2, NDD-T-3, NDD-T-4}. It originates from the dynamical magnetoelectric coupling~\cite{NDD-nagaosa, NDD-enhanced, NDD-skyrmion, ME-electromagnon-1, NDD-ME-1, NDD-ME-2, NDD-ME-3} and has a deep connection with both the order parameter and the geometrical property of Bloch electrons. Per symmetry consideration, NDD requires the breaking of time reversal symmetry~($\mathcal{T}$) and inversion symmetry~($\mathcal{I}$), and is insensitive to the combined $\mathcal{TI}$ symmetry. In three dimensions, such symmetry requirements are generally associated with two types of order parameters, i.e., the multiferroic order $\bm P\times \bm M$~\cite{NDD-nagaosa, NDD-Xray, NDD-enhanced,  NDD-skyrmion, ME-electromagnon-1, NDD-ME-1, NDD-ME-2, NDD-ME-3} ($\bm{P}$ is the electrical polarization and $\bm{M}$ the magnetization) and antiferromagnetism~\cite{ NDD-T-1, NDD-T-2, NDD-T-3, NDD-T-4}. To be precise, in the latter case, the local spin arrangement within a unit cell should possess a spin toroidal order $\bm t\propto \sum_n \bm r_n\times \bm s_n$ or a symmetric spin quadrupole order $q_{ij}^s\propto \sum_n [r_{ni}s_{nj}+r_{nj}s_{ni}-\frac{2}{3}\delta_{ij} \bm r_n\cdot \bm s_n]$~\cite{Spaldin}, where $\bm{r}_n$ and $\bm{s}_n$ label position and spin at lattice site $n$, respectively. Such secondary order parameters then couple with the propagation direction of light, leading to the NDD.
On the microscopic level, based on the band theory analysis in three dimensions, the geometrical origin of NDD is the quantum metric dipole of the Bloch state~\cite{NDD-gao}, which brings unique peak structures to the NDD.

In this Letter, we generalize the microscopic theory of the NDD to two-dimensional antiferromagnets. In sharp contrast, we identify the Berry curvature in the mixed parameter space $(\bm{k},\bm{B})$ as the geometrical origin, instead of the quantum metric dipole. More importantly, in the lattice model of the MnBi$_2$Te$_4$, we find that the Berry curvature exhibits hot regions near the $\Gamma$ point, whose shape is closely tied to the uniaxial strain. Therefore, the magnitude and sign of the NDD can be manipulated by the strain direction, as also dictated by symmetry.

The coupling between the Berry curvature and the uniaxial strain also reveals a distinct mechanism for the NDD. In the presence of strain, $\bm P\times \bm M$ always vanishes due to the combined $\mathcal{TI}$ symmetry; the spin toroidal order $\bm{t}$ is also zero, and the quadrupole order $q_{ij}^s$ is unchanged. Therefore, all of them are irresponsible for the NDD, which strongly depends on the strain.
We then identify the magnetic quadrupole $\mathcal{Q}_{ij}$ of the wave packet as the macroscopic order parameter behind the NDD, whose microscopic expression contains the mixed Berry curvature as well. Utilizing the close relation between the mixed Berry curvature and the strain, we confirm that the NDD depends linearly on $\mathcal{Q}_{ij}$. Therefore, the NDD probes two-dimensional antiferromagnetism through $\mathcal{Q}_{ij}$.

\emph{Geometrical origin of NDD in two dimensions.--} We first sketch the microscopic theory of the NDD in two dimensions. Without loss of generality, we assume that a light linearly polarized along $x$-direction and propagating along $z$-direction, is normally incident on a bilayer antiferromagnet. The multi-layer case can be obtained straightforwardly. The general form of the light-induced current can then be expressed as:
\begin{align}\label{eq_photocurrent}
	J_x(\omega)=\sigma_{xx}(\omega) E_x(\omega)+\sigma_{xxz}(\omega) q_zE_x(\omega)\,,
\end{align}
with $\omega$ and $q_z$ being the frequency and wave vector of the light, respectively. The $\sigma_{xx}(\omega)$ is the usual optical conductivity and $\sigma_{xxz}(\omega)$ is responsible for the NDD, as the resulting current flips sign with the flipping of the light propagation direction, i.e., $q_z$. In three dimensions, $\sigma_{xxz}(\omega)$ can be calculated by considering the energy and momentum transfer during the optical transition. However, the momentum transfer cannot be directly implemented in two dimensions due to the loss of the translational symmetry along the thickness direction.

\begin{figure*}
	\includegraphics[width=18cm,angle=0]{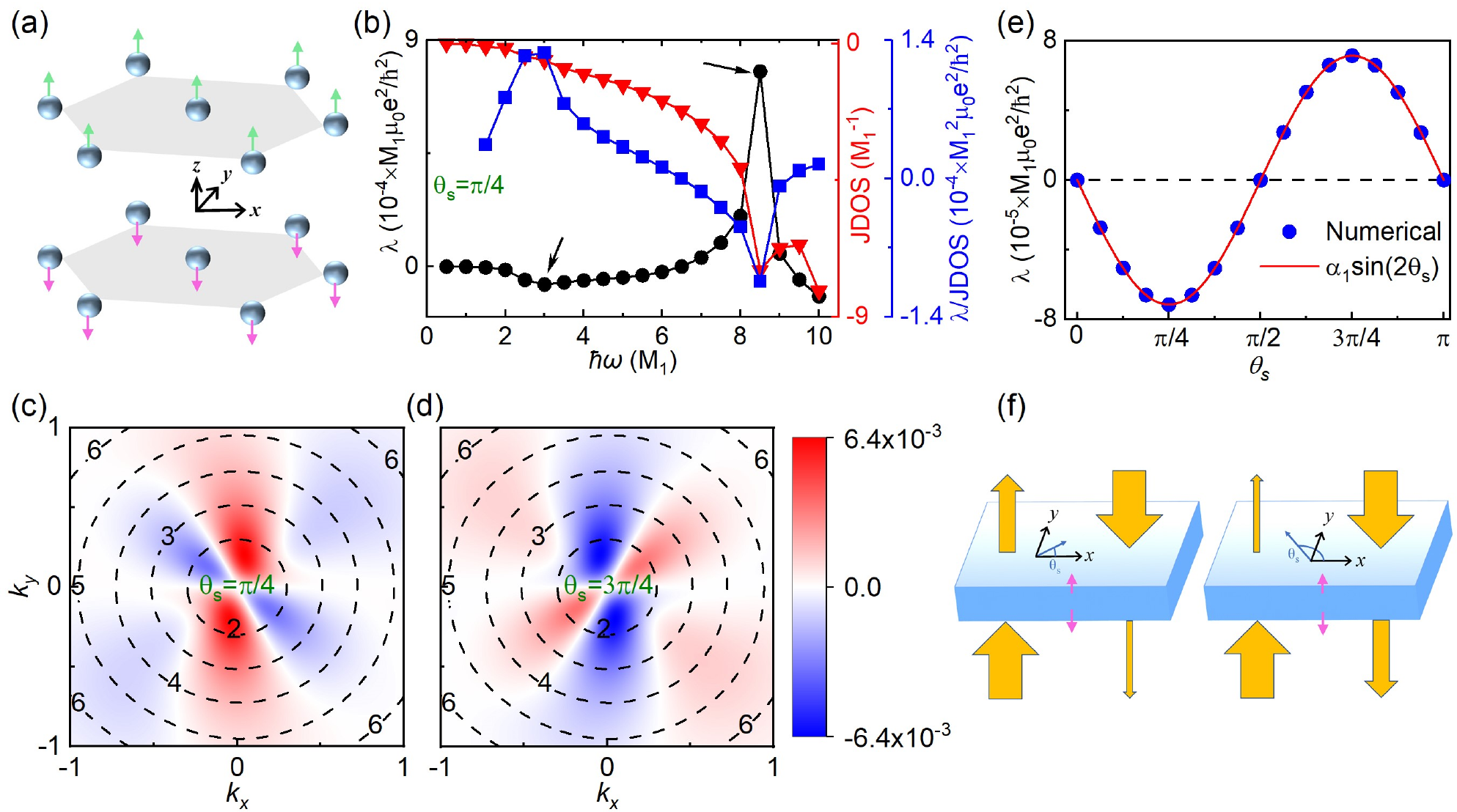}
	\caption{Strain-tunable NDD in bilayer MnBi$_2$Te$_4$. (a) The lattice structure. (b) $\lambda$ ($\mathcal{\lambda}=\frac{\chi}{2d}$), JDOS and $\lambda/{\rm JDOS}$ with a strain angle $\theta_\text{s}=\pi/4$. (c) and (d) The mixed Berry curvature $\Omega_{k_x,\tilde{B}_y}^{mn}$ near the ${\rm \Gamma}$ point with $\theta_\text{s}=\pi/4$ and $3\pi/4$ respectively, where $m$ takes the highest valence band, $n$ takes the lowest conductance band, and the unit is $-de/\hbar$. The dashed curve shows the equal energy contour for optical transition with different photon energy~(in unit of $M_1$).
	(e) The geometrical peak of $\lambda$ as a function of $\theta_\text{s}$, fitted to the function $\alpha_1\sin(2 \theta_\text{s})$.
	(f) A schematic plot of strain-tunable NDD device setup. Pink and yellow arrows represent interlayer antiferromagnetic order, and the incident and transmitted light, respectively.}
	\label{strain_tunable}
\end{figure*}

Following the treatment in Ref.~\cite{stauber-linear response}, we note that the coupling between the electron in the bilayer antiferromagnet and the light vector potential $A_x=A_0 e^{i(q_z z-\omega t)}$($z=\pm d$ for the top and bottom layer) can be put in the form $\hat{H}^\prime=-\hat{ J}_x\cdot A_0 e^{-i\omega t}-2m_y B_y$~\cite{supple}, where $\hat{J}_x=-e \hat{v}_x$ is the current operator, $B_y =i q_z A_0 e^{-i\omega t}$ is the magnetic field sensed by the bilayer antiferromagnet, and $\hat{m}_y=\frac{d}{2}(\hat{J}_{1x}-\hat{J}_{2x})$ is the magnetic moment operator with 2$d$, $\hat{J}_{1x}$ , and $\hat{J}_{2x}$ being the thickness, current operator in the top and bottom layer, respectively. This coupling is given up to first order in both $ A_0$ and $q_z$.

Based on the above light-electron interaction, we use the linear response theory to derive $\sigma_{xxz}(\omega)$. It contains two contributions: the current in response to $B_y$ and the magnetization in response to $E_x$. To make the result compact, we introduce an auxiliary constant magnetic field $\tilde{B}_y$ along $y$-direction, such that the Hamiltonian without light irradiation reads: $\hat{H}=\hat{H}_0-2m_y \tilde{B}_y$, which recovers the original unperturbed Hamiltonian $\hat{H}_0$ in the limit of $\tilde{B}_y\rightarrow 0$. The final result for $\sigma_{xxz}(\omega)$ then reads~\cite{supple}:
\begin{align}\label{eq_sigma}
	\sigma_{x x z}(\omega) = \frac{2 i e }{\hbar \omega} \sum_{m n} \int \frac{d \bm{k}}{(2 \pi)^2} \frac{f_{m n} \varepsilon_{m n }^2}{\varepsilon_{m n }+\hbar \omega+i \eta} \Omega_{k_x,\tilde{B}_y}^{mn}\big |_{\tilde{B}_y\rightarrow 0}\,,
\end{align}
where $f_{mn}=f_{m}-f_{n}$ with $f_{n}$ being the Fermi distribution function for band $n$, $\varepsilon_{mn}=\varepsilon_{m}-\varepsilon_{n}$ with $\varepsilon_{n}$ being the energy dispersion for band $n$, and $\eta\rightarrow 0^+$. $\Omega_{k_x,\tilde{B}_y}^{mn}$ is the Berry curvature in the mixed space $(\bm k, \tilde{\bm B})$ for a pair of bands $m,n$:
\begin{align}
	\Omega_{k_x,\tilde{B}_y}^{mn} = 2 \text{Im}\left(\langle n\bm{k} |\partial_{k_x}| m\bm{k} \rangle \langle m\bm{k} |\partial_{\tilde{B}_y}| n\bm{k} \rangle\right)\,,
\end{align}
where $|n\bm k\rangle$ is the periodic part of the Bloch state in band $n$ for the Hamiltonian $\hat{H}$. 

In Eq.~\eqref{eq_sigma}, the key geometrical quantity is the mixed Berry curvature. To further illuminate its physical and geometrical meaning, several comments are in order. 

Firstly, we emphasize that $\Omega_{k_x,\tilde{B}_y}^{mn}$ is defined for a pair of bands instead of the usual Berry curvature for a single band. This is a characteristic feature in optics as the optical transition always involves a pair of bands. $\Omega_{k_x,\tilde{B}_y}^{mn}$ thus characterizes the geometrical feature of such optical transition. The relation between $\Omega_{k_x,\tilde{B}_y}^{mn}$ and the NDD can be further illuminated by considering the oscillator strength of the optical transition~\cite{dicroich sum-rule}
\begin{align}
	f_{n\rightarrow m}(q_z)=\frac{2m_e}{e\hbar \omega_{mn}}|\langle m\bm k|\hat{J}_x+2iq_z \hat{m}_y|n\bm k\rangle|^2\,.
\end{align}
We then find that~\cite{supple}
\begin{align}
	f_{n\rightarrow m}(q_z)-f_{n\rightarrow m}(-q_z)=\frac{8 q_z m_e \omega_{mn}}{\hbar^2}\Omega_{k_x,\tilde{B}_y}^{mn}\big|_{\tilde{B}_y\rightarrow 0}\,.
\end{align}
Therefore, a nonzero mixed Berry curvature necessarily makes the electron respond differently for counter-propagating lights.

Secondly, $\Omega_{k_x,\tilde{B}_y}^{mn}$ can be viewed as the interband Berry curvature~\cite{niu-rmp}. By summing over the band index $n$ we obtain $\Omega_{k_x,\tilde{B}_y}^m=\sum_n\Omega_{k_x,\tilde{B}_y}^{mn}$. Here $\Omega_{k_x,\tilde{B}_y}^m$ is the familiar Berry curvature in a parameter space $(\bm k,\tilde{\bm B})$ for a single band $m$, and it can be expressed using the Berry connection: $\Omega_{k_x,\tilde{B}_y}^m=\partial_{k_x} A_{\tilde{B}_y}^m-\partial_{\tilde{B}_y} A_{k_x}^m$ with $A_{\xi}^m=\langle m\bm k|i\partial_{\xi}|m\bm k\rangle$ being the Berry connection for band $m$ and $\xi=k_x$ or $\tilde{B}_y$. We comment that $\Omega_{k_x,\tilde{B}_y}^{mn}$ coincides with the familiar Berry curvature if the system only consists of two bands.

The total mixed Berry curvature for the ground state is an intrinsic bulk properties of the Bloch bands and it can be obtained through the optical sum rule. We find that~\cite{supple}
\begin{align}
	\int_{0}^{\infty}\frac{{\rm Re}[\sigma_{xxz}(\omega)] }{\omega} d\omega = 4 \pi e \hbar \sum_{m\in occ} \int \frac{d \bm{k}}{(2 \pi)^2} \Omega_{k_x,\tilde{B}_y}^m\big |_{\tilde{B}_y\rightarrow 0}\,,
\end{align}
where occ means occupied band indices.

Finally, we comment that in experiments the NDD is generally probed by measuring the difference in the transmitted light with the same incident power but opposite propagation directions. The resulting NDD coefficient is usually defined as: $\chi=\frac{\left|\bm{E}_\text{out}(q_z)\right|^2-\left|\bm{E}_{\text {out }}(-q_z)\right|^2}{\left(\left|\bm{E}_\text{out}(q_z)\right|^2+\left|\bm{E}_{\text {out }}(-q_z)\right|^2\right) / 2}$. Using electrodynamic theory, we obtain that~\cite{supple}
\begin{align}\label{eq_chi}
	\chi=-2 \mu_0 \omega \operatorname{Re}\left[\sigma_{x x z}(\omega)\right],
\end{align}
where $\mu_0$ is the permeability of vacuum. Eq.~\eqref{eq_chi} directly relates the NDD to the higher-order optical conductivity $\sigma_{xxz}(\omega)$ and therefore indicates the mixed Berry curvature as its microscopic origin.

\emph{Strain-tunable NDD in bilayer MnBi$_2$Te$_4$.--} To illustrate the geometrical features of NDD, we take bilayer MnBi$_2$Te$_4$ as a concrete example. The lattice structure is shown in Fig.~\ref{strain_tunable}(a) with a magnetic point group $\overline{3}^{\prime} m^{\prime}$~\cite{MBT-symmetry}. The lattice model is given in Ref.~\cite{model}. Although both $\mathcal{T}$ and $\mathcal{I}$ symmetries are broken, which fulfills the spatial temporal symmetry requirement of the NDD, the appearance of $C_{2x}$ symmetry forbids $\sigma_{x x z}(\omega)$ and hence the NDD, based on Eq.~\eqref{eq_photocurrent}. This is also confirmed from our calculation. By applying an in-plane uniaxial strain~\cite{supple}, the $C_{2x}$ symmetry is generally broken and the NDD then emerges. In Fig.~\ref{strain_tunable}(b), we plot the NDD signal per unit thickness~(black curve), defined as $\mathcal{\lambda}=\frac{\chi}{2d}$, with the strain being applied along the direction making a $\theta_\text{s}=\pi/4$ angle with $x$-axis. We find that $\lambda$ has two peaks at $\hbar\omega=3M_1$ and $8.5M_1$, respectively. Here, $M_1$ is a parameter of the lattice model~\cite{supple}.

Optical signals are generally related to the joint density of states defined as,
\begin{align}
\mathrm{JDOS}= \operatorname{Im} \sum_{m\neq n} \int \frac{d \bm{k}}{(2 \pi)^2} \frac{f_{m n}}{\varepsilon_{m n}- \hbar\omega- i \eta},
\end{align}
which represents the number of the initial and final states in the optical transition. By comparing the JDOS~(red curve) and $\lambda$ in Fig.~\ref{strain_tunable}(b), we find that the peak at $\hbar\omega=8.5M_1$ is indeed due to JDOS.

The first peak at $\hbar\omega=3M_1$ is of pure geometrical origin. To understand it, we note that based on Eq.~\eqref{eq_sigma} both the number of states and the mixed Berry curvature contribute to the NDD. We can then isolate the latter by considering the averaged Berry curvature in the optical transition defined as $\lambda$/JDOS, as shown in Fig.~\ref{strain_tunable}(b)~(blue curve). It shows the same two peaks with opposite relative peak strength compared to $\lambda$, clearly showing the essential role of the mixed Berry curvature in the first peak.

To further understand the large peak in $\lambda$/JDOS around $\hbar\omega=3M_1$, we plot the distribution of the Berry curvature near the ${\rm \Gamma}$ point in Fig.~\ref{strain_tunable}(c). The background Berry curvature at zero strain is subtracted as it does not contribute to the NDD signal. We find that the Berry curvature has hot spot area inside $\hbar\omega=3M_1$. Together with the increasing weight factor $\varepsilon_{mn}^2$ in Eq.~\eqref{eq_sigma}, the peak is then shifted toward the blue side around $3M_1$.

Remarkably, the distribution of the mixed Berry curvature is strongly tied to the strain direction. In Fig.~\ref{strain_tunable}(d), we plot the Berry curvature distribution with the strain direction $\theta_\text{s}=3\pi/4$. Comparing Figs.~\ref{strain_tunable}(c) and \ref{strain_tunable}(d), we find that the hot spot area of the Berry curvature distribution changes drastically. Moreover, we note that by a $C_{2x}$ operation, the strain with $\theta_\text{s}=\pi/4$ changes to that with $\theta_\text{s}=3\pi/4$. The mixed Berry curvature $\Omega_{k_x,\tilde{B}_y}^{mn}$ flips sign under $C_{2x}$ due to the $\tilde{B}_y$ derivative. As a result, by flipping the distribution of the Berry curvature according to $C_{2x}$ in Fig.~\ref{strain_tunable}(c) and interchanging the red and blue color, we obtain Fig.~\ref{strain_tunable}(d).

Such a close relation between the Berry curvature and strain leads to a highly tunable NDD. We plot the geometrical peak of $\lambda$ against the strain angle as displayed in Fig.~\ref{strain_tunable}(e). Both the amplitude and sign of NDD signal can be tuned by varying the strain direction. This is in accordance with the $C_{2x}$ symmetry: the strain with $\theta_\text{s}=\theta_0$ is related to that with $\theta_\text{s}=\pi-\theta_0$ and the resulting NDD signal is odd under $C_{2x}$ and hence changes sign; the magnitude of the NDD has to change since at $\theta_\text{s}=0$ and $\pi/2$, the $C_{2x}$ symmetry is restored and the NDD signal vanishes. In fact, $\lambda$ fits the function $\alpha_1\sin(2 \theta_\text{s})$ very well. To further understand it, we note that since $\bm q\times \bm E=\omega \bm B$, the response coefficient $\sigma_{xxz}(\omega)$ is a component of a rank-2 tensor $\beta_{ij}$ with $J_i=\beta_{ij}B_j$ and $\sigma_{xxz}(\omega)=\beta_{xy}/\omega$.  The $C_{2x}$ symmetry forbids $\beta_{xy}$ without strain. By applying strain, $\beta_{ij}$ becomes nonzero and generally has the form of $\beta_{ij}=\gamma_{ijk\ell}S_{k\ell}$ with $S_{k\ell}$ being the strain. The coefficient $\gamma_{ijk\ell}$ is hence a rank-4 tensor, and the point-group symmetry imposes the constraint $\gamma_{ijk\ell}=\det(R) R_{i^\prime i}R_{j^\prime j}R_{k^\prime k}R_{ \ell^\prime \ell} \gamma_{i^\prime j^\prime k^\prime \ell^\prime}$ with $R_{ij}$ being the transformation matrix. For $i=x$ and $j=y$, we find that only two components are nonzero~\cite{supple}, i.e., $\gamma_{xyxy}=\gamma_{xyyx}$. At fixed strain strength $S_0$, the response coefficient has the form
\begin{align}
\beta_{xy}=\beta_0 S_0 \sin(2\theta_\text{s})\,,
\end{align}
where $\beta_0=\gamma_{xyxy}$. This agrees well with the numerical result in Fig.~\ref{strain_tunable}(e).

With the above understanding, the device setup for the NDD in MnBi$_2$Te$_4$ tuned by strain can be illustrated in Fig.~\ref{strain_tunable}(f). By varying the strain direction, we can then choose the favorable propagation direction of light.

\emph{Magnetic quadrupole of the wave packet.--} Surprisingly, NDD in two-dimensional antiferromagnets has a distinct macroscopic origin. In three dimensions, previous experiments have shown that the NDD is a sensitive probe of the multiferroic order $\bm{P} \times \bm{M}$ or the antiferromagnetic order with either a spin toroidal moment $\bm t$ or a spin magnetic quadrupole moment $q_{ij}^s$. However, none of them is responsible for the NDD in bilayer ${\rm MnBi_2Te_4}$ with strain. First, we note that the in-plane uniaxial strain does not break the combined $\mathcal{TI}$ symmetry. Therefore, both $\bm P$ and $\bm M$ are vanishing, and so does $\bm P\times \bm M$. Secondly, under the strain, the lattice sites on different layers move in the same way inside the layer with their distance along the thickness direction unchanged. Since the spin orders on different layers are anti-parallel, one immediately finds that $\bm t$ and the off-diagonal elements of $q_{ij}^s$ are zero, while the diagonal elements of $q_{ij}^s$ are nonzero but unchanged. They cannot explain the sensitive dependence of the NDD on the strain, as discussed previously.

\begin{figure}
	\includegraphics[width=8.5cm,angle=0]{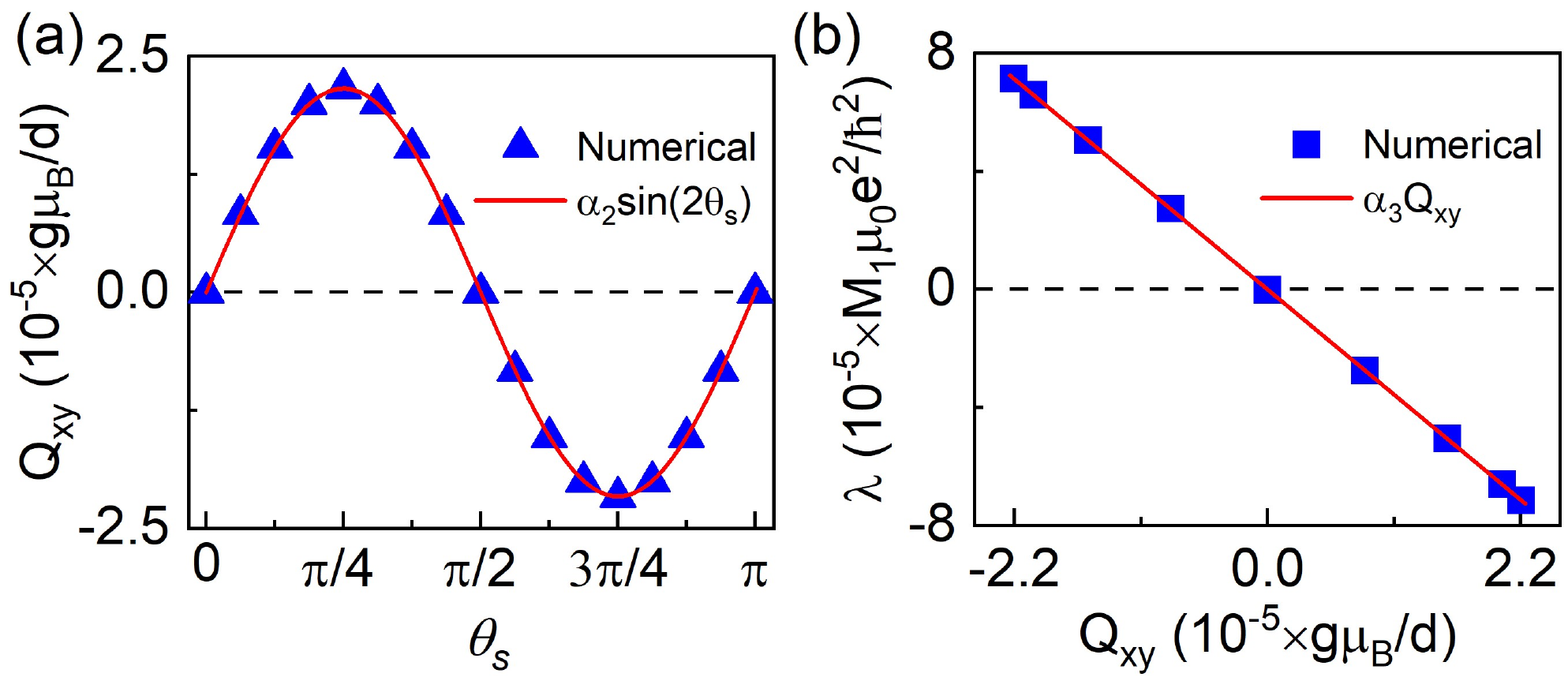}
	\caption{The magnetic quadrupole for the NDD. (a) The strain angle $\theta_\text{s}$ dependence of the magnetic quadrupole $\mathcal{Q}_{xy}$.
	(b) The NDD signal against the magnetic quadrupole $\mathcal{Q}_{xy}$.}
	\label{quadrupole}
\end{figure}

Other than the spin toroidal moment and the spin quadrupole moment due to the spin texture within a unit cell, there is a different definition of spin magnetic quadrupole, labeled as $\mathcal{Q}_{ij}$~\cite{momentumT}, which is for the wave packet of Bloch electrons. It is consistent with the $\mathcal{TI}$ symmetry and is a secondary order parameter associated with the antiferromagnetism as well. Its formal definition reflects the free energy response with respect to a nonuniform magnetic field and its microscopic expression is given by~\cite{supple}
\begin{align}\label{eq_quadrupole}
	\mathcal{Q}_{ij}= & \int_{\text{BZ}} \frac{d\bm{k}}{(2 \pi)^2}\sum_{m\in occ} \bigg\{-\frac{g \mu_\text{B}}{\hbar} \sum_{\substack{n \neq m}}\operatorname{Re}(A^i_{mn} s^j_{nm})\notag\\
	& +(\varepsilon_m-\mu)\Omega_{k_i,\tilde{B}_j}^m\bigg\},
\end{align}
where $g$ is the gyromagnetic factor, $\mu_\text{B}$ is the Bohr magneton, $\mu$ is the chemical potential, $\bm A_{nm}=\left\langle n \bm{k}|i\partial_{\bm{k}}| m \bm{k} \right\rangle$ is the Berry connection, and $\bm s_{m n}=\left\langle m \bm{k}|\hat{\bm{s}}| n \bm{k} \right\rangle$ is the spin matrix element.

Such a magnetic quadrupole can be understood in analogy to the orbital magnetization~\cite{niu-rmp, orbitalm}. We first note that, in Eq.~\eqref{eq_quadrupole}, by changing $-g\mu_\text{B}\hat{\bm s}/\hbar$ to the current operator $-e\hat{\bm v}$ in the first term and $\Omega_{k_i,\tilde{B}_j}^m$ to $-\frac{e}{\hbar}\Omega_{k_i,k_j}^m$ in the second term, we can recover the familiar expression for the orbital magnetization. We can then identify the first term in Eq.~\eqref{eq_quadrupole} as the spin texture within a wave packet, similar to the orbital magnetic moment from the self-rotation of the wave packet in the orbital magnetization. The second term in Eq.~\eqref{eq_quadrupole} is due to the spin polarization on the boundary: there is a confining potential $V$ near the boundary, whose gradient $\bm \nabla V$ yields an electric field; based on the magnetoelectric coupling, such electric field then generates a spin magnetization on the boundary, corresponding to the center-of-mass spin magnetization of the wave packet and hence leading to an additional contribution to the magnetic quadrupole. This physical picture is similar to the boundary current contribution to the orbital magnetization.

The mutual dependence of $\mathcal{Q}_{xy}$ and $\sigma_{xxz}(\omega)$ on the mixed Berry curvature suggests that $\mathcal{Q}_{xy}$ is the macroscopic order parameter sensed by the NDD. This is further confirmed by utilizing the close relation between the mixed Berry curvature and the uniaxial strain. We first confirm that at zero strain, $\mathcal{Q}_{xy}$ vanishes identically, just as the NDD. In Fig.~\ref{quadrupole}(a), we plot $\mathcal{Q}_{xy}$ against the strain angles. We find that it exhibits a sinusoidal profile similar to the NDD in Fig.~\ref{strain_tunable}(e). By directly plotting the NDD against $\mathcal{Q}_{xy}$ at different strain angles as shown in Fig.~\ref{quadrupole}(b), we find that the NDD is indeed linearly correlated to $\mathcal{Q}_{xy}$.

In summary, we identify the mixed Berry curvature as the geometrical origin of the NDD for two-dimensional antiferromagnets. It is strongly tied to the uniaxial strain, offering a highly tunable NDD. Such a close relation between the Berry curvature and the strain also reveals the magnetic quadrupole of the wave packet as the macroscopic order parameter for the NDD in two dimensions, which is distinct from the multiferroic order $\bm P\times \bm M$ or the spin toroidal and quadrupole order in previous studies.

Y. Gao is supported by the National Key R\&D Program (Grant Nos. 2022YFA1403502) and Fundamental Research Funds for the Central Universities (Grant No. WK2340000102). W. Liang and Z. Qiao are supported by the National Natural Science Foundation of China (Grant Nos. 11974327), Fundamental Research Funds for the Central Universities (WK3510000010, and WK2030020032), Anhui Initiative in Quantum Information Technologies (AHY170000), and Innovation Program for Quantum Science and Technology (2021ZD0302800). J. Zeng is supported by the National Natural Science Foundation of China (No. 12247181). Q. Niu is supported by the National Natural Science Foundation of China (No. 12234017). We also thank the Supercomputing Center of University of Science and Technology of China for providing the high performance computing resources.

\end{document}